\documentclass[12pt]{iopart}
\pdfoutput=1

\usepackage{graphicx}
\usepackage{dcolumn}
\usepackage{bm}
\usepackage{soul}
\usepackage{color}

\begin{document}

\title[Manipulating the orientation of an organic adsorbate on silicon]{Manipulating the orientation of an organic adsorbate on silicon: a NEXAFS study of acetophenone on Si(001)}

\author{Kane M O'Donnell$^{1}$\footnote{Present address: Department of Imaging and Applied Physics, Curtin University, Bentley WA 6102, Australia}, Oliver Warschkow$^{2}$, Asif Suleman$^{3,4}$, Adam Fahy$^{5}$, Lars Thomsen$^{1}$ and Steven R Schofield$^{3,4}$}
 \address{$^{1}$Australian Synchrotron, Clayton, VIC 3168, Australia.}
\address{$^{2}$Centre for Quantum Computation and Communication Technology, School of Physics, University of Sydney, Sydney NSW 2006, Australia.}
\address{$^{3}$London Centre for Nanotechnology, University College London, London WC1H 0AH, United Kingdom.}
\address{$^{4}$Department of Physics and Astronomy, University College London, London WC1E 6BT, United Kingdom.}
\address{$^{5}$School of Mathematical and Physical Sciences, University of Newcastle, Callaghan NSW 2308, Australia.}

\ead{steven.schofield@physics.org}

\date{\today}

\begin{abstract}
We investigate the chemical and structural configuration of acetophenone on Si(001) using synchrotron radiation core-level spectroscopy techniques and density functional theory calculations. Samples were prepared by vapour phase dosing of clean Si(001) surfaces with acetophenone in ultrahigh vacuum. Near edge X-ray adsorption fine structure spectroscopy and photoelectron spectroscopy measurements were made at room temperature as a function of coverage density and post-deposition anneal temperature. We show that the dominant room temperature adsorption structure lies flat on the substrate, while moderate thermal annealing induces the breaking of Si-C bonds between the phenyl ring and the surface resulting in the reorientation of the adsorbate into an upright configuration.
\end{abstract}

\maketitle

\section{\label{sec1}Introduction}

There is currently a strong interest in the creation of novel nano- and molecular-scale structured materials and devices.  This is motivated both by a desire to further develop our understanding of condensed matter physics at these length scales, and by the potential for the development of the next generation of (opto)electronic devices.  
The direct covalent attachment of organic molecules to silicon surfaces has long been considered a route toward  the incorporation of molecular functionality with conventional semiconductor technology\cite{Wolkow1999}.  
Recently, the molecular functionalization of silicon surfaces has been used for applications in conventional electronics\cite{ITRS2013}, such as the creation of molecular CMOS (complementary metal oxide semiconductor) devices\cite{Cummings2011}, hybrid molecule-silicon memory capacitor structures\cite{Wakayama2014,Pro2009}, and molecular sensors\cite{Vilan2002}.  Molecular electronic junctions consisting of individual molecules attached to silicon surfaces and contacted by a metallic gate have been proposed and measured\cite{Yaffe2009,Vilan2010}, with recent observations including current rectification\cite{Haj-Yahia2012}.  This progress toward functional silicon-organic devices is extremely encouraging and in parallel there has been considerable progress in the development of the fundamental understanding of molecule-semiconductor interactions\cite{Coustel2013,TaoBook2012,Harikumar2011,Bent2011}.  Nevertheless, much work remains to be done, particularly in the  development of methods to control adsorbate structural and electronic properties; i.e., methods for attaching molecules in a desired structural (e.g., standing up versus lying down) and electronic (strongly/weakly electronically coupled) configurations.  We note further that these properties also strongly affect junction properties such as interface dipole formation that can have a dominant effect on the device characteristics\cite{Yaffe2009,Vilan2002}.

To this end we have studied an interesting model system: the interaction of acetophenone with the technologically ubiquitous Si(001) surface.  Acetophenone has two functional groups, a phenyl ring and an acetyl group, each of which can in-principle form covalent bonds with the silicon surface.  However, for molecular electronics applications it is desirable for the molecule to attach to the surface such that its electrically conductive $\pi$-conjugated phenyl ring is unperturbed by direct bonding to the substrate.  
The adsorption of acetophenone to clean Si(001) surface has been studied using scanning tunnelling microscopy/spectroscopy (STM/STS) and density functional theory (DFT)\cite{Schofield:2013gf}. It was proposed that acetophenone first adsorbs datively via the carbonyl oxygen atom and subsequently undergoes a sequence of reaction steps that lead to an allyl-radical (figure 1a). In this configuration, which is stable at room temperature, the adsorbate lies flat on the substrate.  Direct stimulation of an individual adsorbate using the STM tip, or moderate thermal annealing of the sample, was shown to convert some of the adsorbed molecules into an alternative structure that imaged brightly in STM. It was proposed that this feature results from a structure where the phenyl ring is no longer bound directly to the substrate but is instead oriented almost perpendicular to the plane of the surface (figure 1b).

\begin{figure}
\includegraphics{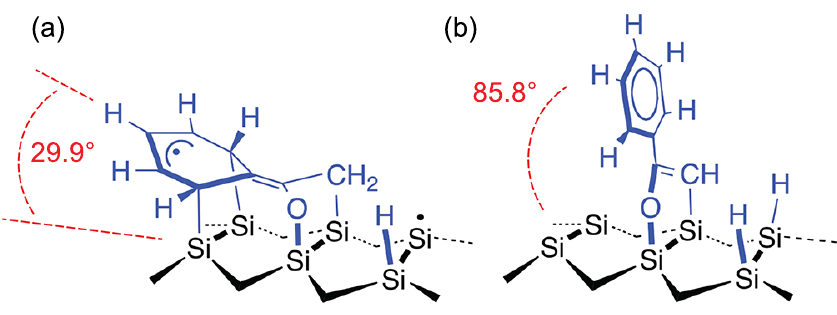}
\caption{\label{figure1} Schematic valence and orthographic structure views of the two main configurations of acetophenone on Si(001). Throughout the text, positions in the phenyl ring are numbered 1-6 clockwise from the acetyl group relative to these structures. (a) The allyl-radical configuration in which the phenyl ring is bonded flat to the surface at ring carbon atoms 2 and 6, with $\pi$ conjugation across carbons 3, 4, and 5 as indicated. The $\pi$ system makes an angle of $29.9^\circ$ with the surface plane. (b) The ``tall feature'' configuration in which the phenyl ring is not directly bound directly to the substrate, and the $\pi$ conjugated phenyl ring makes an angle of $85.8^\circ$ with the surface plane. This small tilt angle with respect to the surface normal is attributed to the asymmetry between the oxygen and methyl bonding sites and the steric hindrance of the hydrogen attached to the methyl side.}
\end{figure}

Here, we use near-edge X-ray absorption fine structure (NEXAFS) spectroscopy, and X-ray photoelectron spectroscopy (XPS) to measure the orientation of $\pi$-conjugated molecular orbitals of acetophenone when adsorbed to Si(001). NEXAFS is a synchrotron spectroscopy technique that is sensitive to both the chemical environment of the atomic absorption sites and the orientation of the unoccupied molecular orbitals in the vicinity of the core hole, providing a means to accurately determine the orientation of molecules adsorbed on surfaces\cite{Stohr:2003vl}. Our NEXAFS measurements, which we perform as a function of both surface coverage and substrate temperature, provide strong evidence that acetophenone initially adsorbs in a configuration parallel to the surface upon dosing at room temperature.  Upon moderate thermal annealing, we observe a reversal in the dichroism of the NEXAFS spectra that is characteristic of a structural rearrangement of the adsorbates into an upright configuration. This interpretation is reinforced by XPS data that shows the Si-C bonding is reduced subsequent to annealing, which agrees with the proposed structure where the phenyl ring is not directly bonded to the surface.  These measurements are in agreement with the structural configurations proposed in \cite{Schofield:2013gf} (Figs. 1a and 1b), and provide an important independent confirmation of the reorientation of acetophenone on Si(001) subsequent to moderate thermal annealing. In addition, we present measurements performed over a wide range of surface coverages, from submonolayer to saturation, going considerably beyond the earlier STM/DFT work \cite{Schofield:2013gf} that considered only very low coverage. We support our spectroscopic measurements with density functional theory calculations that give direct insight into the origin of spectral features in the two structural configurations.

\section{\label{sec2}Experimental Methods}

Experiments were conducted at the Soft X-Ray beamline of the Australian Synchrotron\cite{Cowie:2010cm}. Both NEXAFS and XPS measurements were carried out at room temperature under ultra high vacuum (UHV) conditions with a base pressure of better than $2.0 \times 10^{-10}$ mbar. Silicon samples were cleaved from $0.04 - 0.06$ Ohm cm (001) orientated silicon wafers (Virginia Semiconductor Inc.) and degassed in UHV by direct current heating to $\sim 600^{\circ}$C overnight. Before XPS/NEXAFS measurements, and/or exposure of the sample to acetophenone, the sample was flash annealed to $\sim 1200^{\circ}$C for 10 s with a controlled ramp-down over 2 minutes. The silicon samples were allowed to cool for 30 minutes before dosing with acetophenone, which was done from vapour using a precision leak valve with the dose flux calculated using the total chamber pressure. Acetophenone (99.0\%, Sigma Aldrich) was purified by several freeze/pump/thaw cycles and checked for the absence of residual water contamination by UHV residual gas analysis (Stanford Research Systems SRS100). For each dose, a series of 10 minute direct-current anneals were used to investigate changes in the adsorbed species with temperature. The direct current heating power was recorded and used as an indirect measurement of the anneal temperature. Anneals were carried out in steps of 0.1 A from 0.2 to 0.7 A; at the highest anneal temperatures both XPS and NEXAFS indicated damage to the molecules and the formation of silicon carbide-like structures. 
Following each sample preparation our measurement procedure was as follows: the sample was transferred into the analysis chamber and a gold 4f XPS energy calibration spectrum was taken from a gold reference foil mounted directly beneath the sample. Carbon 1s and silicon 2p XPS spectra were recorded immediately following this Au energy reference measurement. A photon energy of 330 eV was used to gain a maximal carbon cross section of the C 1s peak. Carbon NEXAFS spectra were then recorded (275 to 320 eV) at five angles between normal (90$^{\circ}$) and grazing (20$^{\circ}$) photon incidence. Partial electron yield spectra\cite{Stohr:2003vl} were collected by recording the Auger electron yield from the sample with the repelling grid set to 230 V. Normalization was performed using the Òclean monitor methodÓ as described by Watts \etal\cite{Watts:2006gi}, by measuring a clean Si(001) surface and monitoring the flux signal via a 50\% transmissive gold grid placed upstream of the sample along the beamline. After the NEXAFS spectra were acquired, further XPS spectra for Au 4f, C 1s and Si 2p were collected to monitor for any changes due to contamination or beam damage.

NEXAFS data was analyzed in the usual way following the procedures described in \cite{Stohr:2003vl, Watts:2006gi}: all spectra were processed by a constant-value pre-edge subtraction and post-edge normalization to unity. The spectra were then fitted using Gaussian peaks and an error step function with an exponential decay. The purpose of NEXAFS spectroscopy in this work is to measure spectral dichroism, that is, the change of peak intensity with the incident beam angle, which probes the orientation of the adsorbates on the surface\cite{Stohr:2003vl}. In order to measure these orientations, the peak area of the $\pi^\star$ resonance was plotted against the incident angle $\theta$ of the photon beam and fitted using the following equation (see Eq. 9.23 in \cite{Stohr:2003vl}):

\begin{equation}
\label{equation1}
I(\theta) = I_{0}\left(\cos^{2}{\theta}\left(1 - \frac{3}{2}\sin^{2}{\alpha}\right) + \frac{1}{2}\sin^{2}{\alpha}\right)
\end{equation}

Equation \ref{equation1} is valid when the adsorbed molecule has three-fold or higher in-plane symmetry either intrinsically or due to the symmetry of the substrate. In the present case, due to the underlying symmetry of the dimerized terraces, there is four-fold symmetry for each possible molecular state. The prefactor $I_0$ is a free fitting parameter for a given NEXAFS resonance and $\alpha$ is the tilt angle of the resonance orbital with respect to the surface. In using this form of Equation \ref{equation1}, the polarization of X-ray light has been assumed to be unity, an excellent approximation for an undulator insertion device as used on the Soft X-Ray beamline of the Australian Synchrotron\cite{Cowie:2010cm}. By fitting the $\pi^{\star}$ transition to Equation \ref{equation1} we can deduce the angle of the plane of the transition with respect to the substrate surface.

XPS spectra were fitted using Voigt functions after linear pre-edge and Shirley background subtraction\cite{Shirley:1972ts}.  The Voigt shape parameter was constrained to 0.3 to fix the Gaussian/Lorentzian character, with resulting Lorentzian widths in the range 0.13-0.18 eV and Gaussian widths in the range 0.35-0.50 eV.  

\section{\label{section3}Computational Methods}

In order to assign the spectral features of our NEXAFS experiments to molecular features, we calculated
theoretical NEXAFS spectra using the plane wave/pseudopotential density functional theory code CASTEP\cite{Clark:2005vp}. 
 In these calculations, the Si(001) surface was represented using a five-layer slab geometry with a dihydride termination at the 
bottom surface. The top surface to which the adsorbate molecules are attached represents 
two dimer rows of four dimers each; i.e.~we used a (4$\times$4) surface unit cell which affords an adequate degree of 
separation between the adsorbate and its in-plane periodic images. In the surface-perpendicular direction, 
a period repeat of 25~{\AA} was used. This results in a vacuum separation between successive slabs of approximately 
18~{\AA} (surface to surface) and at least 11~{\AA} between the top of an adsorbed molecule and dihydride termination of the next slab.
The Perdew-Burke-Ernzerhof (PBE)\cite{Perdew:1996ur} density functional was used to describe electron exchange and correlation, and ultrasoft pseudopotentials\cite{Vanderbilt:1990uj} were used to represent the contribution of the core electrons. The plane wave basis set was truncated using a cutoff energy of 490~eV and the irreducible Brillouin zone was sampled at the $\Gamma$ point only. All atoms of the adsorbate molecule and the silicon slab were fully relaxed to a force convergence criterion of 0.02~eV/{\AA}. The positions of the hydrogen atoms of the dihydride termination were held fixed in order to emulate the strain due to the deeper silicon bulk that is not explicitly represented in our model.

Theoretical carbon $1s$ NEXAFS spectra were computed via the following method. A `core hole' pseudopotential was generated for carbon where half an electron was removed from from the 1s state during pseudopotential generation. The resulting pseudopotential was used to represent the (X-ray) absorbing atom in a single-point calculation. For each carbon atom in a given slab, an independent, self-consistent field calculation was carried out to generate both the excited state eigenvalue spectrum and the transition matrix elements between the carbon 1s core state and each excited state. Of course, because CASTEP is a pseudopotential code, the Kohn-Sham states resulting from the self-consistent field procedure are in fact pseudowavefunctions, not the all-electron states. Hence, as detailed by Gao, \etal\cite{Gao:2009fk} a projector augmented wave (PAW) reconstruction is used within the CASTEP code to regenerate the all-electron wavefunctions prior to matrix element calculation. This is accomplished by using the ultrasoft projectors from the pseudopotential itself to act as a basis for the all-electron wavefunction within the atomic sphere of the absorbing atom and taking advantage of the close correspondence between the ultrasoft pseudopotential method and the PAW formalism. The initial core state $|i\rangle$ used to generate the matrix elements is obtained from an atomic all-electron calculation. The matrix elements generated using this PAW reconstruction show little deviation from those generated using an all-electron code\cite{Gao:2009fk}.

The set of transition matrix elements from the core state to excited states allows the calculation of the absorption cross section $\sigma_{xn}$ for the $n$th carbon absorbing site via Fermi's golden rule:\cite{Gao:2009fk, Mizoguchi:2004fr, Mizoguchi:2009gd, Tanaka:2005hq}

\begin{equation}
\label{equation2}
\sigma_{xn} \propto \sum_{f} |\langle f | \varepsilon \cdot \nabla | i \rangle |^{2} \delta(E_{f} - E_{i} - \hbar\omega)
\end{equation}

In this equation $E_f$ and $E_i$ are the Kohn-Sham energies for each unoccupied state $f$ and the initial state $i$ and $\varepsilon$ is the electric field vector. The result is a set of absorption cross sections that must be summed together to get the total absorption cross section for the molecular adsorbate. However, considerable care is required dealing with the energy scale in Equation \ref{equation2}. First, the atomic all-electron core state eigenvalue is not suitable as the initial state eigenvalue in Equation \ref{equation2} because as is well known from XPS, the core level energies are altered by the local chemical environment of the atom. Hence the delta function is shifted onto a scale where $E_i = 0$ and the lowest unoccupied state also has zero energy. This does not affect the shape or intensity of the resulting spectrum. Then, to put all the spectra from each atomic site on a common energy scale, a transition energy was calculated for each separate cross section $\sigma_{xn}$ using the method of Mizoguchi, et al\cite{Mizoguchi:2009gd}. Briefly, this method uses the difference in total energies between the excited system for a particular atomic absorber and that of a ground-state calculation with no core-hole pseudopotential. However, extra terms are required to avoid double-counting certain components of the energy due to the use of pseudopotentials. These extra energy terms are obtained from isolated atomic calculations as detailed by Mizoguchi et al\cite{Mizoguchi:2009gd}. Each individual spectrum for a given atomic absorber within the molecule was then shifted by the corresponding transition energy before the individual spectra were added together. The spectra were then broadened first using a Lorentzian function with linearly increasing width to replicate core-hole lifetime effects and the photon linewidth of the X-ray source, and finally a Gaussian function to replicate instrument broadening and thermal effects. 

Theoretical tilt angles were determined by first calculating the spectra for varying photon incidence angles, followed by a fit of the calculated peak areas to Eq. \ref{equation1}. This procedure mirrors that used for the experimental NEXAFS spectra.

\section{\label{section4}Results}

We performed detailed measurements at three different acetophenone coverages by dosing with 0.05, 0.23 and 2.3 L, which we refer to as the ÒlowÓ, ÒintermediateÓ, and ÒhighÓ dose regimes, respectively.  The low dose was chosen to be as close as possible to our detection limit where the signal to noise ratio in the spectra is poor.  This low coverage regime is of interest since it most closely matches the conditions measured in the STM experiments in \cite{Schofield:2013gf} where the adsorbed molecules are isolated from one another.  The intermediate dose (0.23 L) was chosen such that we improve the signal to noise of the data, but are still within a regime where the adsorbates are not significantly affecting one another.  We then increased the dose by a further order of magnitude (2.3 L) to produce a high coverage of adsorbates on the surface. Significant spectral differences in both XPS and NEXAFS spectra confirm that this high dose corresponds to a saturation, or near-saturation, coverage.

Figure 2 shows two sets of angle-dependent NEXAFS spectra of the intermediate coverage surface: Fig. 2(a) is of the surface after dosing, while Fig. 2(b) is of the surface after dosing with a subsequent thermal anneal by direct current heating (0.3 A, 2.7 W for 10 minutes). Both sets of spectra clearly exhibit features that are characteristic of molecular adsorption. In particular there are sharp resonances at $284.6 \pm 0.1$ and $286.0 \pm 0.1$ eV, which we label A and B, respectively, followed by broader resonances and a step in the total intensity at higher photon energy at around 291.5 eV. The sharp resonances are transitions to $\pi^{\star}$ molecular orbitals, while the broad resonances at higher energies are transitions to $\sigma^{\star}$ orbitals. In particular, we identify a very broad $\sigma^{\star}$ resonance with a width of about 10 eV centred around 298 eV that we label (C) in figure 2(a). 

\begin{figure}
\includegraphics{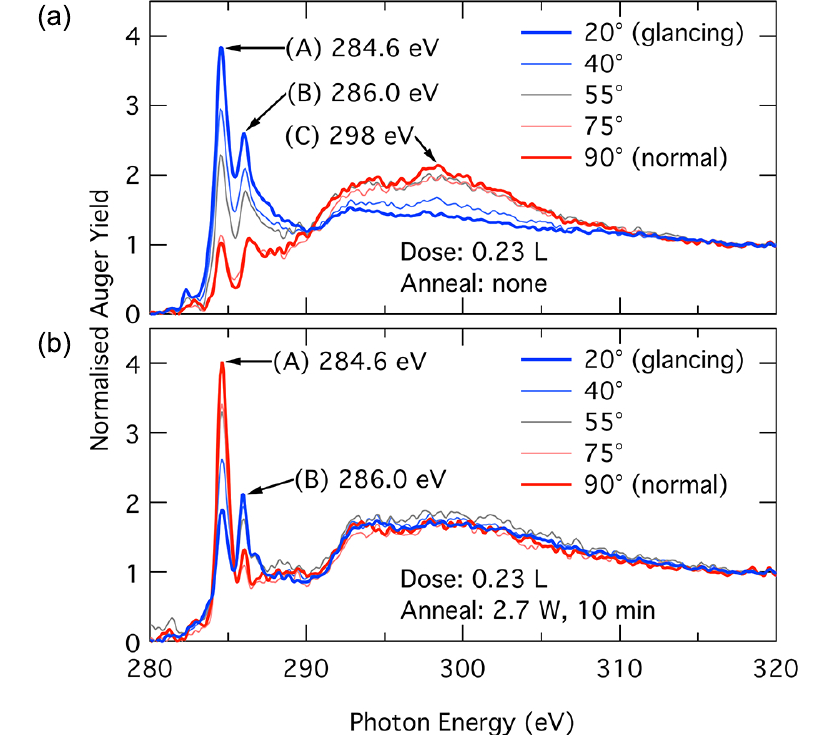}
\caption{\label{figure2} Angle-dependent NEXAFS spectra of Si(001) exposed to 0.23 L of acetophenone at room temperature, which we classify as an intermediate level dose. (a) Spectra acquired directly after dosing.  (b) A sample dosed under identical conditions as in (a) and then subjected to a direct current anneal at 2.7 W for 10 minutes. Glancing ($20^\circ$) and normal incidence ($90^\circ$) spectra are highlighted using blue and red coloured lines, respectively.}
\end{figure}

Comparing the spectra of the intermediate dosed sample with and without the anneal step [Figs. 2(b) and 2(a), respectively], we note that dramatic changes occur in the dichroism of the molecular peaks. Most notably, peak A undergoes a dichroism reversal: i.e., peak A is intense in the glancing angle  (20$^{\circ}$) spectrum and weak in the normal angle (90$^{\circ}$) spectrum for the dosed surface, but this reverses in the annealed sample. Peak B does not undergo a dichroism reversal, but its intensity variation with angle is considerably reduced in the annealed sample. There is no detectable dichroism of the broad $\sigma^{\star}$ resonance in the annealed sample. 

In figure 3 we present a repeat of the experiments shown in figure 2 but for the low coverage surface (0.05 L). The signal to noise ratio is considerably worse in this data, reflecting a much lower quantity of carbon on the surface.  It is nevertheless evident that these spectra contain the same features (peaks A, B, and C) in similar proportion to the intermediate coverage data shown in figure 2 for both the dosed and annealed samples. It is notable that the dichroism of peak A reverses after dosing, and the dichroism in peak C is visible in the dosed-sample but absent in the annealed. These qualitative observations confirm both the reproducibility of our measurements and the interpretation presented above that these two doses correspond to sub-monolayer adsorbate coverages where the adsorbates are essentially not interacting with one another.

\begin{figure}
\includegraphics{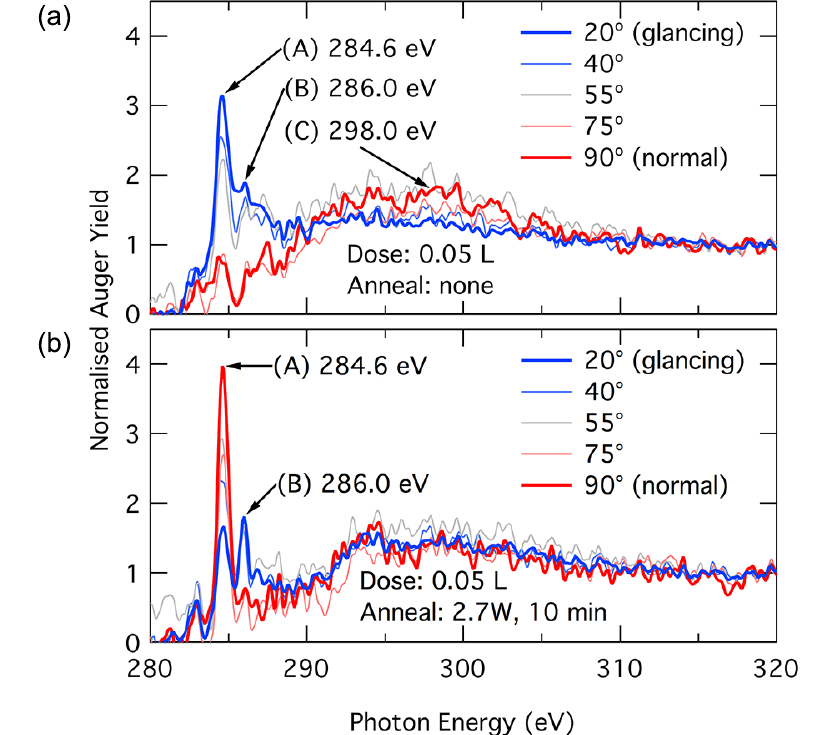}
\caption{\label{figure3} Angle-dependent NEXAFS spectra of Si(001) exposed to a low dose (0.05 L) of acetophenone at room temperature: (a) dosed sample, and (b) dosed and direct current annealed to 2.7 W for 10 minutes.}
\end{figure}

Figure 4 presents the spectra for the high-dose surface (2.3L). Note that this high coverage surface was annealed using a higher direct current heating power, 4.1 W, than for the two lower coverage surfaces. These data show the $\pi^{\star}$ and $\sigma^{\star}$ peaks at the same energies as the lower coverages, and exhibit qualitatively similar trends in their variation in the ratio of peak intensities.  However, subtle differences suggest that the coverage is high enough that the configuration(s) of the adsorbates on the surface is not entirely the same as for the lower coverages. In particular, the intensity ratio of peaks A and B has changed significantly in the high coverage spectra. 

\begin{figure}
\includegraphics{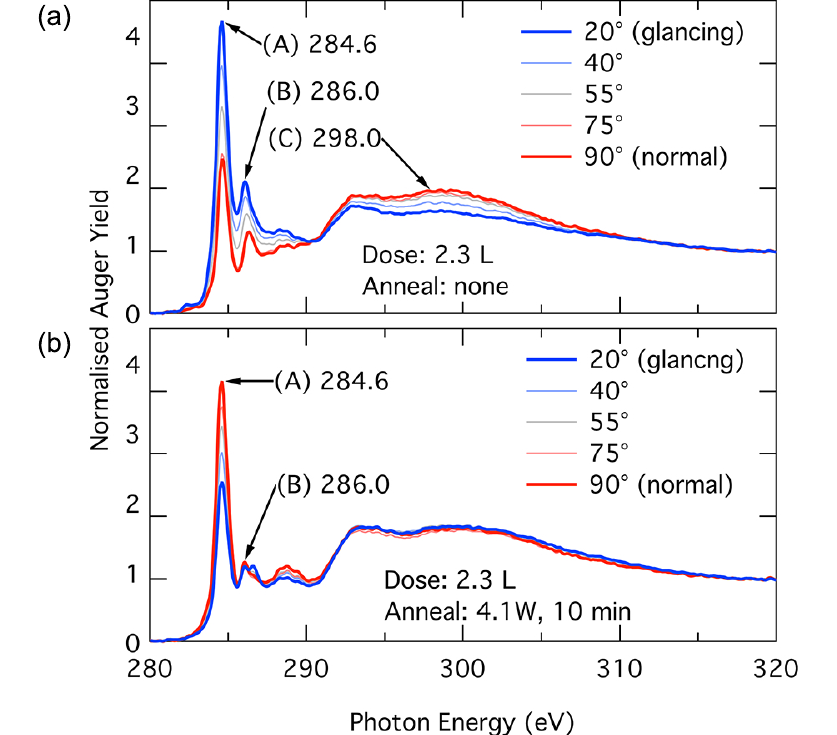}
\caption{\label{figure4} Angle-dependent NEXAFS spectra of Si(001) exposed to a high (2.3L) dose of acetophenone at room temperature. (a) dosed sample, and (b) dosed and direct current annealed to 4.1 W for 10 minutes.}
\end{figure}

We next examine quantitatively the variation of the NEXAFS peak A intensity before and after annealing for each of the three coverage regimes. Figure 5 shows the variation of peak A as a function of the substrate angle with respect to the beam for the three doses, with, and without annealing. By fitting to Eq. \ref{equation1} as described in the Methods section, for the low and intermediate doses we deduce orbital tilt angles of $37 \pm 13^{\circ}$ and $36 \pm 3^{\circ}$ respectively for the dosed surface, rising to $67 \pm 3^\circ$ and $68 \pm 3^\circ$ for the annealed surfaces. For the high dose the equivalent values are $45 \pm 4^\circ$ and $63 \pm 3^\circ$ respectively. Error estimates are given as one standard deviation derived from the diagonal elements of the fit covariance matrix for each data set.

\begin{figure*}
\includegraphics[width=\textwidth]{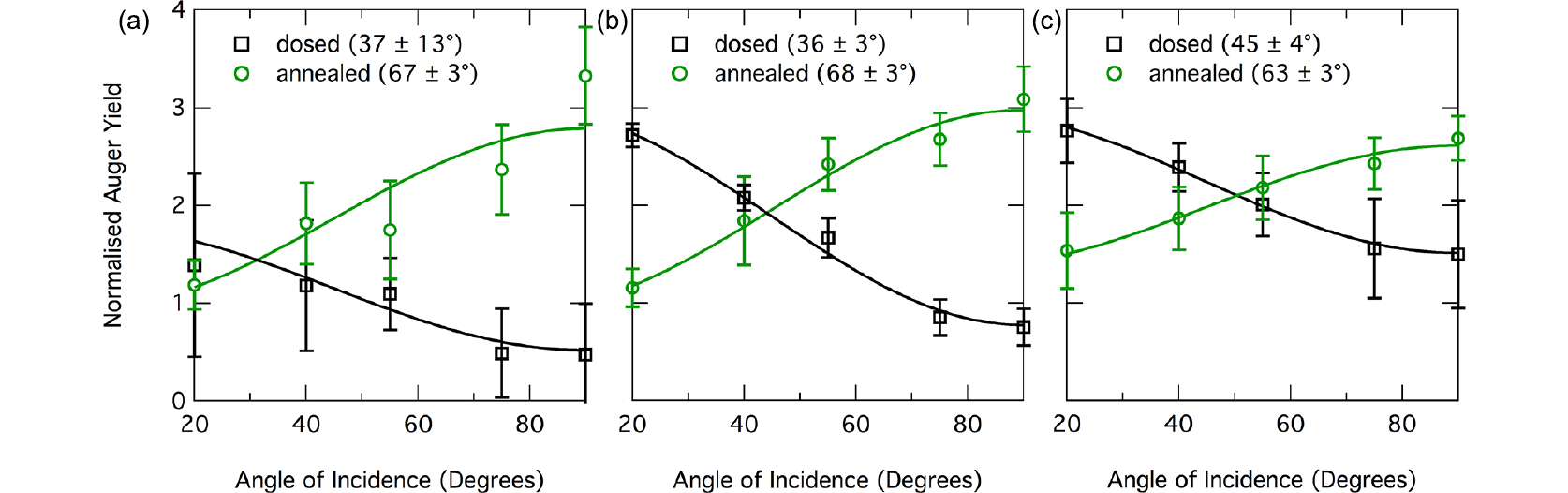}
\caption{\label{figure5} Angular variation of the intensity of peak A (284.6 eV) in the NEXAFS spectra shown in Figs. 2-4, i.e., for silicon (001) exposed to a (a) low (0.05 L), (b) medium (0.23 L), and (c) high (2.3 L) dose of acetophenone at room temperature both with and without a subsequent thermal anneal. The intensities and their uncertainties were determined from Gaussian peak fitting to the NEXAFS spectra and are fitted with Eq. \ref{equation1}, as described in Methods.}
\end{figure*}

We have simulated NEXAFS spectra for the $\pi^{\star}$ region based on DFT calculations of the two structures shown in figure 1.  The simulated spectra for the allyl-radical and tall feature are shown in Figs. 6a and 6b, respectively, and we note the strong agreement with the experimental spectra for the low and intermediate coverages presented in Figs. 2 and 3. With the theoretical spectra shifted so that peak A is aligned with experiment, we find that peak B is 0.8 eV higher in the theoretical spectra for both the allyl radical and tall feature. This difference from experiment, less than 3\% of the experimental transition energy, is consistent with other DFT studies of absorption spectra\cite{Mizoguchi:2009gd, Taillefumier:2002de, Gao:2008fb} and reflects the well-known limitations of DFT with respect to the energetic determination of excited states\cite{Godby:1986ud}. We simulate grazing and normal incidence theoretical spectra to show the angular dichroism expected for each structure. Accompanying the theoretical spectra are orbital densities for the lowest unoccupied state of the allyl radical (figures 6c, 6d) and the tall feature (figure 6e, 6f) to illustrate the shape of the $\pi^{\star}$ final state for the dominant transition of peak A. In allyl-radical (figure 6c, 6d), the absorbing (excited) atom is on position 3 of the phenyl ring, within the 3, 4, 5 conjugated system of the adsorbate. In the tall feature (figure 6e, 6f), the absorbing atom is position 4 of the phenyl ring. 

\begin{figure}
\includegraphics{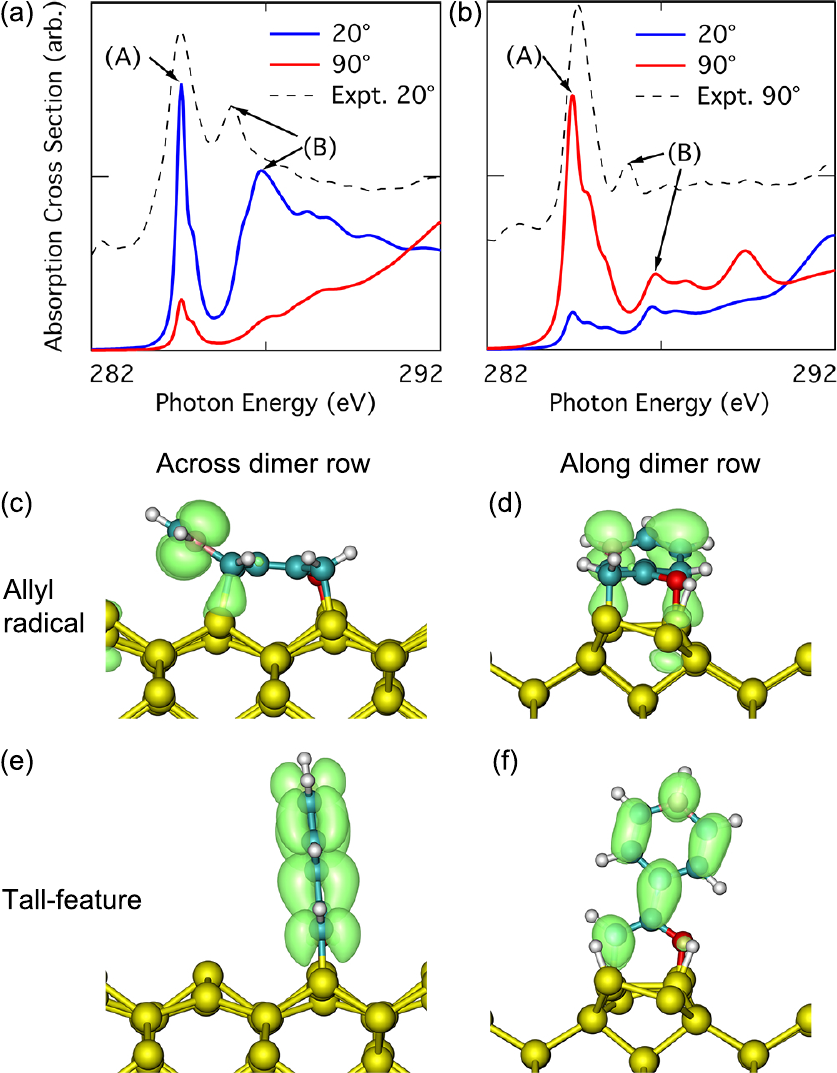}
\caption{\label{figure6} Simulated NEXAFS spectra for two adsorption structures of acetophenone on Si(001): (a) The allyl-radical structure, (b) the tall feature. For comparison, two experimental NEXAFS spectra from our intermediate dosed surface (figure 2) have been included as dashed lines.  The energy axis of the simulated spectra has been set such that feature A is aligned to that of the experimental spectrum. (c, d) Orbital density plots for the lowest unoccupied state of the allyl radical, with an excited atom at position 3 of the phenyl ring; the two views shown are looking across and along the surface dimer row, respectively.  (e, f) Orbital density plots for the lowest unoccupied state of the tall feature with an excited atom at position 4 of the phenyl ring, looking across and along the dimer rows, respectively.}
\end{figure}

To gain additional information on the surface adsorbate structures we also recorded XPS C $1s$ spectra before and after each set of NEXAFS measurements.  A comparison of the before and after spectra for each sample confirms negligible time dependent changes, ruling out any deleterious beam damage or contamination issues. However, as might be expected, there are significant changes in the XPS spectra due to substrate annealing. Figure 7 shows the C $1s$ spectra for the intermediate dose (figure 7a) and after annealing (0.3A, 2.7 W; figure 7b). The dosed surface is dominated by a broad peak that is fit by two components at 284.2 eV and 284.5 eV, with a higher-energy feature at 286.3 eV well-separated from the main peak. After annealing, the primary peak is best fit by a single component at 284.6 eV with a new low binding energy feature at 283.4 eV and a shift in the higher-energy feature to 285.9 eV. All components have a position uncertainty of $\pm$ 0.1 eV.

\begin{figure}
\includegraphics{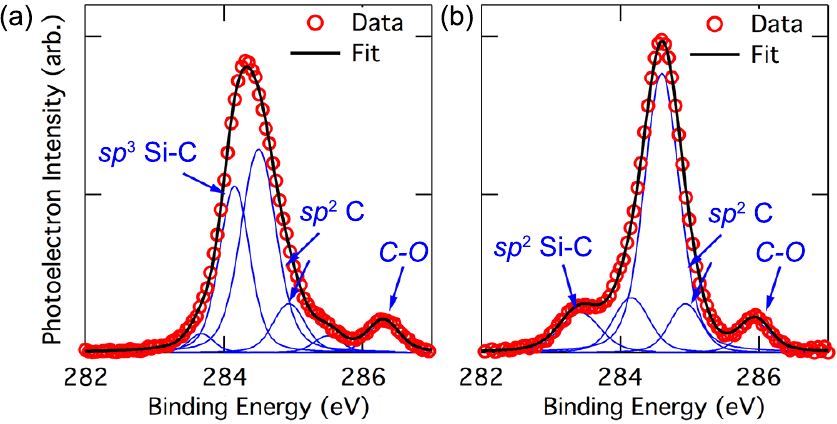}
\caption{\label{figure7} Carbon $1s$ core level XPS spectra of silicon (001) exposed to a flux of 0.23 L (our intermediate level dose) of acetophenone at room temperature, (a) before and (b) after annealing. Each spectrum is fit using a combination of Voigt functions as described in Methods.  Suggested origins for the spectral components are indicated and agree well with the allyl radical and tall feature adsorption structures, as described in the text.}
\end{figure}

\section{\label{section5}Discussion}

The dichroism seen in the NEXAFS spectra results from the Òsearchlight effectÓ, where resonance intensities are strongest when the electric field vector of the x-ray photons is parallel to the lobes of the final state\cite{Stohr:2003vl}.  Samples with highly ordered molecular states therefore give rise to strongly dichroic spectral variations with incident photon beam angle.  

Peak A at 284.6 eV can be identified as a transition to $\pi^{\star}$ orbitals of an aromatic or conjugated system (e.g. see \cite{Solomon:1990jn, Solomon:1991wk, Coulman:1991wm, Yang:1995ug, Watts:2011kk}). For both the low and intermediate coverage surfaces, our data thus demonstrates a predominance of an adsorption structure incorporating $\pi^{\star}$ orbital states that make an angle of $\sim 36^\circ$ with respect to the surface. This is in good agreement with the predicted allyl-radical structure, where the primary aromatic/conjugated component is made up of positions 3, 4 and 5 in the phenyl ring; our DFT calculations of this structure indicate that the local tilt angle of this system is $29.9^\circ$ (figure 1a), which is in quite good agreement with our measured values of $37  \pm 12^\circ$ and $36 \pm 3^\circ$.

In the case of the annealed surface we measure an angle of $68 \pm 3^\circ$, which prima facie does not appear to agree with the structure presented in figure 1b, where the phenyl ring is tilted at $85.8^\circ$ relative to the surface (figure 1b). We believe that this discrepancy results from the fact that the annealed acetophenone surfaces do not consist of a single adsorbate configuration, but rather they exhibit a mix of configurations. The NEXAFS spectra from a surface containing such a mixture of structures are then linear combinations of spectra from each configuration\cite{Urquhart:1999vc}. Since only two features were observed in the STM experiments of annealed acetophenone dosed Si(001), and no stable structures with intermediate tilt angles ($\sim 30^\circ < \theta < \sim85^\circ$) are known\cite{Schofield:2013gf}, we propose that the tilt angles measured in our low and intermediate coverage NEXAFS measurements correspond to a linear combination of the allyl-radical and tall features illustrated in figure 1. Based on this assumption we can conclude that the annealed surface we measure in the low and intermediate coverage regimes corresponds roughly to $32 \pm 6$\% of horizontally aligned allyl-radical structures and $68 \pm 6$\% upright tall features. 
  
Our simulated NEXAFS spectra (figure 6) and their excellent agreement with the measured spectra (Figs. 2 - 4) strongly support the interpretation presented above and provide additional insights into the origin of the individual spectral components. Peak A in the simulated NEXAFS spectra for the allyl-radical (figure 6a) is dominated by contributions from positions 3 and 5 in the phenyl ring (figure 6c). Applying Eq. \ref{equation1} to the intensity variation of the theoretically derived peak A produces a tilt angle of $30^\circ$, as expected. Feature B in the theoretical spectra is primarily made up of contributions from position 1 of the phenyl ring and the carbonyl site on the acetyl group, and corresponds well to the experimental peak B for the dosed surface. The theoretical spectra for the tall feature is shown in figure 6b. The primary peak (A) in these spectra corresponds to $\pi^{\star}$ contributions from the phenyl ring (figure 6d), and applying Eq. \ref{equation1} we obtain a tilt angle for the phenyl ring of $82.6^\circ$. If we assume that peak A in the annealed spectrum in figure 2b is a combination of peaks from both the allyl-radical and tall feature structures we find that a allyl radical to tall feature ratio of 0.28:0.72 ($\pm 0.06$) reproduces the experimentally measured tilt angle of $68 \pm 3^\circ$. In other words, $72 \pm 6$\% of the acetophenone has been converted into the vertically-aligned tall features via annealing, in excellent agreement with the estimate above of 68\% based on the bond angles in the structural models. Peak B in the spectrum for the tall feature has the same origin as for the allyl radical (i.e. position 1 of the phenyl ring), but is of much weaker relative intensity. This explains why we only see dichroism reduction and not reversal for peak B after annealing: peak B in the annealed spectra is dominated by the fraction of the adsorbed molecules remaining in the allyl radical configuration.

Annealing the low and intermediate coverage acetophenone dosed surfaces at slightly lower heating power (0.2 A, 1.8 W; not shown), produced a much-reduced dichroism for both peaks A and B.  This is consistent with a more equal mix of the two surface structures as would be expected by a thermally activated process and a lower anneal temperature.  Furthermore, annealing at higher temperatures (0.4 A, 4.1 W to 0.8 A, 8.4 W; not shown) also reduced the observed peak A and B dichroism from that in Figs. 2 and 3, eventually removing it altogether which we interpret as the thermal decomposition of the adsorbates toward the expected eventual formation of a SiC surface\cite{Palermo:2006et}.  Further measurements with a finer mesh of annealing points are required in order to determine whether a surface consisting of only tall features can be thermally prepared. 

For the high coverage surface we find a measured tilt angle of $45 \pm 4^\circ$, which is significantly closer to the so-called ``magic angle'' of 57.4$^\circ$ for NEXAFS where one cannot distinguish between a particular tilt angle and random orbital orientations\cite{Stohr:2003vl}. This is consistent with the interpretation presented above that in this system the adsorbates are interacting with each other resulting in a more complex arrangement of surface adsorbate structures after dosing.  For example, some fraction of the adsorbates may be sterically hindered from onwards reaction by the dative bonded configuration or other intermediate structural configurations that are not thermally stable at room temperature for adsorbates on the clean surface. After annealing, the measured orbital tilt angle for peak A is close to that measured at low coverages ($63 \pm 3^\circ$); however, we note that the measured dichroism maximum occurred at a higher annealing power compared to the low and intermediate doses (4.1 W versus 2.7 W), which is also consistent with the interpretation of a saturated surface with interacting adsorbate configurations. 

We now make some final remarks on our XPS measurements and how they further confirm the interpretation of our NEXAFS data presented above. The XPS spectrum from the dosed surface exhibits a broad primary peak that must be fit with at least two components at 284.2 and 284.5 eV. After annealing, this main peak is narrowed and can be fit by a single peak at 284.6 eV; the 284.2 eV component is seen to substantially reduce. Rochet, \etal, studied the adsorption of ethylene on Si(001)\cite{Rochet:1998wk}, a system where all the carbon atoms in the molecule are directly bonded to silicon atoms in a $sp^3$ configuration. They found the dominant C 1s component at $284.2 \pm 0.2$ eV. On this basis we assign the 284.2 eV component of the primary peak to $sp^3$-like Si-C bonding. For the allyl radical, this corresponds to the methyl carbon plus positions 2 and 6 of the phenyl ring.  The reduction in this component with annealing is therefore consistent with the breaking of Si-C bonds between the phenyl ring and the silicon surface that is expected from the conversion from allyl-radical to tall feature. For the dosed surface we associate the 284.5 eV component with the 3, 4, 5 conjugated system of the allyl radical. After annealing, all the phenyl carbons become equivalent in the aromatic ring yielding the 284.6 eV dominant feature in that spectrum. In addition, another feature becomes apparent at 283.4 eV; this can be attributed to the methyl carbon in the acetyl group, which in the tall feature is bonded to silicon with $sp^2$ character after the further hydrogen dissociation. The other noticeable change in the C $1s$ spectrum is the shift in the high binding energy feature that is attributed to carbon bonded to oxygen. This feature moves to lower binding energies after annealing, suggesting a slight change in the charge density in the vicinity of the acetyl group as might be expected moving from the allyl radical converting to the doubly dissociated tall feature.  

\section{\label{section6}Conclusions}

In this work we have used high-resolution carbon NEXAFS to measure the orientation of ¹ conjugated orbitals of acetophenone adsorbates on the Si(001) surface. We find that surfaces with submonolayer coverages of acetophenone produced by ``low'' and ``intermediate'' acetophenone flux exhibit $\pi^{\star}$ tilt angles of $37 \pm 13^\circ$ and $36 \pm 3^\circ$, respectively. These measurements are consistent with an ``allyl-radical'' adsorbate structure, which is predicted by DFT to have a $\pi$ conjugated system at an angle of $30^\circ$ with the surface plane. After moderate thermal annealing, the measured tilt angles increase to $67 \pm 3^\circ$ and $68 \pm 3^\circ$, respectively. The larger tilt angles unambiguously demonstrate that some fraction of the adsorbates have undergone a structural rearrangement, and we interpret our measurements in light of an acetophenone adsorption structure where the phenyl ring is upright and the molecule is bonded to the surface only via its acetyl group. We infer that approximately 70\% of the allyl-radical adsorbates have converted to this upright configuration as a result of annealing.  We use DFT calculations to produce simulated NEXAFS spectra, which reproduce the experimental spectra and provide insight into the origin of the individual spectral components. XPS C $1s$ spectra measured with and without substrate annealing demonstrate a reduction in the Si-C bonding and other spectral differences further confirming the interpretation of our NEXAFS data. At higher acetophenone flux we measure a NEXAFS tilt angle of $45 \pm 4^\circ$, which suggests the presence of additional adsorption structures with different molecular orientations, as may be expected for a saturation dosed surface. Annealing the high coverage surfaces resulted in an increase in the measured tilt angle to $63 \pm 3^\circ$, consistent with the lower coverage measurements although a higher anneal temperature was required. A careful study of the annealing process may lead to a method for a uniform functionalization of covalently bonded, functional molecules with a preserved electronic conjugation to the surface.

\ack
This research was undertaken on the Soft X-Ray beamline at the Australian Synchrotron, Victoria, Australia.The authors are grateful for the technical support of B. Cowie and A. Tadich of the Australian Synchrotron. We acknowledge financial support from the Engineering and Physical Sciences Research Council (EP/H003991/1; EP/L002140/1), and the Australian Synchrotron.  O.W. is supported by the Australian Research Council Centre of Excellence for Quantum Computation and Communication Technology (project number CE1100096). A. F. acknowledges a University of Newcastle postgraduate scholarship. This work used computational resources provided by the Australian National Computational Infrastructure (NCI) and the multi-modal Australian sciences imaging and visualisation environment (MASSIVE). 

\section*{References}
\bibliography{ReferencesJun2014}

\end{document}